\documentclass[lettersize,journal]{IEEEtran}

\usepackage{amsmath,amsfonts}
\usepackage{algorithmic}
\usepackage{array}
\usepackage[caption=false,font=normalsize,labelfont=sf,textfont=sf]{subfig}
\usepackage{textcomp}
\usepackage{stfloats}
\usepackage{url}
\usepackage{verbatim}
\usepackage{cite}
\usepackage{siunitx}
\usepackage{mathtools}
\usepackage{breqn}

\usepackage{graphicx}
\graphicspath{{./figures/}}

\def\BibTeX{{\rm B\kern-.05em{\sc i\kern-.025em b}\kern-.08em T\kern-.1667em\lower.7ex\hbox{E}\kern-.125emX}}

\usepackage{balance}

\usepackage{hyperref}
\usepackage{booktabs}
\usepackage{diagbox}
\usepackage{multirow}
\usepackage{tablefootnote}

\usepackage{makecell}

\DeclareMathOperator{\sigm}{sigm}

\usepackage{color}

\begin{document}

\title{Transportation mode recognition based on low-rate acceleration and
  location signals with an attention-based multiple-instance learning network}

\author{Christos Siargkas, Vasileios Papapanagiotou,Anastasios Delopoulos

  \thanks{Christos Siargkas and Anastasios Delopoulos are with the Multimedia
    Understanding Group, Dpt. of Electrical and Computer Engineering, Faculty of
    Engineering, Aristotle University of Thessaloniki, Greece.
    
    Vasileios Papapanagiotou is with the IMPACT research group, Dpt. of
    Biosciences and Nutrition, Karolinska Institutet, Stockholm, Sweden and the
    Multimedia Understanding Group, Dpt. of Electrical and Computer Engineering,
    Faculty of Engineering, Aristotle University of Thessaloniki, Greece.
    
    E-mail: \url{csiargka@ece.auth.gr}, \url{vasileios.papapanagiotou@ki.se}, \url{antelopo@ece.auth.gr}
    
    Manuscript received August 21, 2023; revised February 13, 2024. This
    research was funded by the European Union's H2020 programmes under Grant
    Agreement No. 965231 (REBECCA - ``REsearch on BrEast Cancer induced chronic
    conditions supported by Causal Analysis of multi-source data'').}}

\markboth{Siargkas et al.: TMR based on low-rate acceleration and location
  signals with an attention-based MIL network}{IEEE Transactions on Intelligent
  Transportation Systems, vol. XX, no. X, XXXXXXXX 2024}

\maketitle

\begin{abstract}
  Transportation mode recognition (TMR) is a critical component of human
  activity recognition (HAR) that focuses on understanding and identifying how
  people move within transportation systems. It is commonly based on leveraging
  inertial, location, or both types of signals, captured by modern smartphone
  devices. Each type has benefits (such as increased effectiveness) and
  drawbacks (such as increased battery consumption) depending on the
  transportation mode (TM). Combining the two types is challenging as they
  exhibit significant differences such as very different sampling rates. This
  paper focuses on the TMR task and proposes an approach for combining the two
  types of signals in an effective and robust classifier. Our network includes
  two sub-networks for processing acceleration and location signals separately,
  using different window sizes for each signal. The two sub-networks are
  designed to also embed the two types of signals into the same space so that we
  can then apply an attention-based multiple-instance learning classifier to
  recognize TM. We use very low sampling rates for both signal types to reduce
  battery consumption. We evaluate the proposed methodology on a publicly
  available dataset and compare against other well known algorithms.
\end{abstract}

\begin{IEEEkeywords}
  transportation mode detection, accelerometer, location, GPS, multiple instance
  learning, attention, hidden Markov model
\end{IEEEkeywords}

\section{Introduction}
\label{sec:introduction}

To discover solutions and uncover opportunities to improve the quality of life
at scale in cities, it would be useful to acquire a better understanding of
commuting patterns and establish new grounds of collaboration between the
transportation sector and other fields of research like biomedical research
\cite{Straczkiewicz2019, Cooper2000, Loprinzi2013}, urban and transportation
planning \cite{Kaplan2012}, public-policy making, environmental research
\cite{Froehlich2009, Brazil2013}, carbon footprint analysis \cite{oroeco}, safe
driving, and journey planning.  Transportation mode recognition (TMR) systems
can play an integral part in this by automating the process and eliminating the
need of manual data collection and annotation. Moreover, real-time recognition
can assist in identifying crucial moments and providing immediate assistance.

Given the rich arsenal of sensors available in common smartphone devices (i.e.,
smartphones and smartwatches), data fusion has become an integral part of TMR
research, by combining multiple sensors to observe the same event from different
viewpoints.

Acceleration signal can elaborately distinguish between physical activities
which are characterized by volatile body movement, however, the ability to
distinguish between motorized transportation modes (TMs) that are characterized
by low body movement is lesser \cite{Tang2023, Wang2019}. In particular, when
using only acceleration signal, classifiers can recognize activities such as
standing still, walking, and running robustly, but exhibit significant
ambiguities between car vs. bus, train vs. subway, and standing still
vs. train/subway. Findings of \cite{Tang2023} align with these limitations,
highlighting the model’s constrained ability to differentiate between Subway and
Train.

Contrary, location signals can capture a ``higher-level'' overview of the user's
movement, including distance, speed, etc; however, relying location signals
requires uninterrupted interaction with satellites which is not always possible
(i.e., when the user moves underground, inside a building, or along urban
canyons). The inherent heterogeneity between these two sensor modalities
presents an opportunity for highly-effective fusion, yet it also introduces a
challenge for achieving seamless combination. Furthermore, the vastly different
sampling rates used for collecting the two sensor modalities introduce another
layer of complexity in robustly combining them.

In this work, we propose an intermediate-fusion deep-learning TMR model which
combines two sensor modalities: acceleration and location. To address the
challenges posed by variations in these two sensor modalities, including
heterogeneity, distinct sampling rates, and sensor unavailability, we propose a
novel model that learns independent spatio-temporal features for each modality
while simultaneously mapping the heterogeneous sensor data to a shared
low-dimensional embedding space. Leveraging this common embedding space, we
efficiently integrate information from the diverse sensors using an
attention-based multiple-instance-learning (MIL) classifier. Finally, we employ
a Hidden Markov Model (HMM) for post-processing. To optimize the efficacy of
MIL, we use sequences of acceleration windows instead of relying on single
windows. This approach offers several advantages, such as enhancing the
resolution of the acceleration input and identifying the most critical regions
within the acceleration data that pertain to the specific TM being analyzed. We
evaluate on a publicly available dataset and compare with our own
implementations of other state-of-the-art algorithms.

Our main contributions are:
\begin{itemize}
\item We propose a novel, energy-efficient, and lightweight system for TMR that
  uniquely combines low sampling-rate acceleration and location signals
  collected by smartphone sensors, optimizing energy consumption while
  maintaining high effectiveness.
\item We present Fusion-MIL, an attention-based MIL framework that effectively
  combines acceleration and location signals, overcoming challenges such as
  sensor heterogeneity, distinct sampling rates and sensor unavailability. We
  use multiple acceleration windows instead of single one, enhancing the
  resolution of the acceleration input and identifying regions most relevant to
  the final prediction. This approach inherits the interpretability of the
  instance-attention based MIL.
\item We introduce additional processes such as data pre-processing, feature
  engineering, data augmentation, and pre-training to boost effectiveness.
\item Extensive experimental evaluation is performed on a publicly available
  dataset, with a focus on cross-subject and cross-placement variability. These
  experiments provide in-depth insights into various factors influencing TMR,
  particularly the impact of device placement. Results demonstrate our method's
  capability to accurately distinguish between eight different TMs in complex
  scenarios, surpassing the state-of-the-art methods and various alternative
  single-modal and multi-modal algorithms.
\end{itemize}

The rest of the paper is organized as follows: Section \ref{sec:related}
presents an overview of recent approaches for TMR based on inertial signals,
location signals, or their combination. Section \ref{sec:materials} presents our
proposed approach and the architecture of the artificial neural network
(ANN). Section \ref{sec:setup} presents the dataset, types of experiments, and
details for model training, and Section \ref{sec:results} presents the results
and experimental evaluation.  Finally, Section \ref{sec:conclusions} concludes
the paper.

\section{Related Work}
\label{sec:related}

Inertial and location signals offer different advantages and challenges when
used for transportation mode recognition (TMR). In this section we present
recent approaches for TMR from literature, organized based on their use of
sensor type, with an emphasis on the requirements (i.e., sampling rate, battery
consumption).

\subsection{Location-based approaches}

One of the most commonly used modalities for TMR is the location signal (i.e.,
geographical coordinates). In \cite{James2021}, James et al. explore this domain
using the publicly available GeoLife Dataset \cite{zheng2011geolife} and process
location trajectories (sampling interval of $1$ to $5$ sec) based on discrete
wavelet transform (DWT) and deep learning techniques. In a different approach,
Dabiri el al. \cite{Dabiri2018a} adjust all the location trajectories to a fixed
length and design a CNN-based supervised travel mode identification system. This
system extracts $4$ features from location data, including speed, acceleration,
jerk, and bearing and identifies $5$ types of TMs.

It is important to note that these studies, as well as the vast majority of the
TMR studies, use the same set of users for training and testing, limiting the
generalization potential to user variations such as different gait styles,
walking/running patterns, speed, and travel behaviours.

In \cite{Stenneth2011}, location-based TMR using a random forest (RF) is
combined with geographic information system (GIS) to leverage the underlying
transportation network. Location is sampled only once every $15$ seconds to
avoid battery drain. The trained classification model is deployed to the general
public, accommodating new individuals for evaluation. Authors report an average
detection accuracy of $93.5\%$ at classifying $6$ TMs: car, bus, train, walking,
biking and stationary. The 2021 SHL Recognition Challenge \cite{SHL2021} also
focuses on user-independent evaluation by separating participants in train and
test sets (instead of separating segments of participant data). The contestant
teams were provided with four radio sensor modalities, including: GPS location,
GPS reception, WiFi reception, and Cell reception. These sensors were
asynchronously sampled with a sampling rate of roughly $1$ Hz. Sekiguchi et
al. \cite{Sekiguchi2021}, using an LSTM model with location-derived features
(velocity, acceleration, and angular acceleration) and XGBoost with statistical
features for when location signal was not available achieved the 5th highest
F1-score of $58.2\%$. In \cite{Balabka2021}, Balabka and Shkliarenko extract
$975$ hand-picked features from the $4$ sensors and employ an AdaNet model,
provided by Google AutoML service, taking the first place with F1 score as high
as $75.4\%$.

\subsection{Inertial based approaches}

Location signal is not always available (especially when it is derived from GPS)
in shielded areas; additionally, sampling-rates can quickly drain the smartphone
battery. Thus, several studies propose only inertial-based only approaches
\cite{Alaoui2021}. Most of them only rely on the accelerometer due to its low
energy requirements and early integration into smartphones.

In \cite{Liang2019}, 3D acceleration signals are collected from $4$ participants
holding their phones freely in any orientation of their preference. Acceleration
signal is captured at $50$ Hz. After filtering and computing the signal
magnitude, a CNN model discriminates between $7$ TMs: stationary, walk, bicycle,
car, bus, subway and train. This approach achieves an accuracy of $94.48\%$ by
splitting the data into $80\%$ - $20\%$ training and test sets
respectively. Nonetheless, like several other studies, this study employs the
same group of users for both training and testing. Contrary, Rosario et
al. \cite{Rosario2014} observe that when a classifier is deployed on data of an
older participant after being trained on a younger participant, its
effectiveness considerably declines.

In \cite{Hemminki2013}, $3$ common body placements are considered and
acceleration is recorded both at $60$ Hz and $ 100$ Hz. A hierarchical
classifier is proposed to identify $6$ TMs. This approach is evaluated using a
leave-one-subject-out (LOSO) cross-validation (CV) and a leave-one-placement-out
CV, achieving F1-scores of $81\%$ and $83.3\%$ respectively.

However, some vehicles such as automobiles and buses (or subways and trains)
produce similar patterns, thus limiting the discriminative power of
acceleration-only based systems \cite{Chen2017}. To counter this, some studies
propose combining multiple inertial sensor modalities.

Fang et al. \cite{Fang2017} uses $8,311$ hours of accelerometer, magnetometer,
and gyroscope sensor data recorded from $224$ users with a sampling rate of $30$
Hz \cite{Yu2014}. A deep neural network (DNN) is employed to recognize $5$ TMs
(still, walk, run, bike, and vehicle), achieving accuracy of $95\%$. Yet, a key
challenge in TMR, especially when relying solely on inertial sensors, is the
discrimination ability between similar motorized classes \cite{Nikoli2017,
  Gjoreski2018}.

In Zhao et al. \cite{Zhao2019}, a Bi-LSTM model is trained on accelerometer and
gyroscope data, sampled at $50$ Hz, to detect $6$ TMs: still,
walk, run, bike, bus, subway. The model is trained on data from $8$ users and
is evaluated on $3$ separate users, achieving an overall accuracy of
$92.8\%$. However, all data is collected from smartphones with a fixed
orientation and placement, limiting generalization to other body positions. In
fact, according to \cite{Guvensan2018, Yang2016}, inertial  data may differ
significantly across different body positions. In \cite{Tang2023}, Tang et
al. use a deep multimodal fusion network on accelerometer, magnetometer, and
gyroscope  data sampled at $20$ Hz and segmented into $60$-second
windows. Their proposed method achieves a remarkable accuracy of $95.1\%$ and an
F1-score of $94.7\%$ in detecting eight different transportation modes.

The 2018 SHL recognition challenge \cite{SHL2018} focuses on time-invariant
evaluation using $62$ days ($272$ hours) as training data and the remaining $20$
days ($95$ hours) as test data, while providing the participant teams with $7$
different motion-sensor modalities. Ito et al. \cite{Ito2018} apply a deep CNN
to two-dimensional spectrogram images extracted from the sensor sequences,
taking the third place with F1-score $88.8\%$. In the position-independent 2019
SHL recognition challenge \cite{SHL2019}, Choi and Lee \cite{Choi2019a} achieve
the second highest F1-score $75.9\%$ by introducing a deep multi-modal fusion
model named EmbraceNet that combines multiple inertial sensor modalities which
are independently pre-processed by a CNN.

\subsection{Hybrid approaches}

Compared to inertial sensors, location sensors can offer a ``higher'' overview
of the user's movement, thus serving as a complementary data source when used
alongside inertial sensors. Moreover, location sensors are mostly position
invariant and can be effectively combined with inertial sensors to minimize the
impact of the varying smartphone placement and orientation. Consequently, a
combination of location and inertial sensor modalities can enhance the
robustness of a TMR system and improve its discrimination abilities. In fact,
according to Wang et al. \cite{Wang2019}, the combination of location and
accelerometer outperforms the combination of three inertial sensor modalities
(accelerometer, gyroscope, and magnetometer), while the combination of all four
modalities only improves the recognition effectiveness slightly. Additionally,
\cite{Wang2019} also emphasize that the combination of location and acceleration
signals remarkably improves the recognition accuracy for each class compared to
relying solely on accelerometer data. Notably, the accuracy of distinguishing
between Car and Bus, as well as between Train and Subway, shows significant
improvement with the combined use of location and accelerometer data which is
what we propose in this work.

In \cite{reddy2010}, Reddy et al. introduces a hybrid approach using both
location and acceleration data captured from $6$ different smartphones (worn in
$6$ different body positions) by $16$ participants. Each smartphone contains an
accelerometer that can sample at $32$ Hz and a built-in GPS receiver that can
sample at $1$ Hz. The proposed classifier consists of a DT followed by a
discrete hidden Markov model (HMM) and recognizes $5$ TMs: still, walk, run,
bicycle, motorized. The classifier is evaluated with LOSO CV, achieving an
average and a minimum accuracy of $93.6$\% and $88.2$\% respectively. However,
this method can't handle location-signal loss.

In \cite{Widhalm2012}, location (GPS) trajectories are resampled evenly with a
frequency of $1$ Hz and the accelerometer readings are resampled to a frequency
of $50$ Hz. The proposed approach copes with location signal loss by including
positioning data obtained from the cell network and relying solely on
accelerometer features when the trajectory cannot be reconstructed with
sufficient accuracy. A randomized ensemble of classifiers with an HMM is
proposed to differentiate among $8$ different TMs: bus, car, bike, tram, train,
subway, walk, and motorcycle. Two experiments are presented: a $4$-class task
(walk, car, bus, bike) where the classifier achieves a high recall accuracy of
$91\%$, and a $5$-class task (walk, car, bus, subway, train) where
classification effectiveness is degraded (recall accuracy of $76\%$).

In \cite{Priscoli2020}, Priscoli et al. use accelerometer ($50$ Hz), gyroscope
($50$ Hz), and location ($1$ Hz) signals to recognize seven TMs: walk, car,
motorbike, tram, still, bus, and subway. This is the only work, in our
knowledge, that proposes a deep learning TMR system based on hybrid information
from both location and inertial sensor modalities. Specifically, recursive
neural networks (RNNs) with statistical features achieve the second best
effectiveness ($88\%$), while a CNN model with 3D acceleration, 3D gyroscope,
and speed achieves $98.6\%$ without any pre-processing.

In \cite{Giri2022}, location, accelerometer, and heart rate data are collected
from $126$ individuals over a period of $7$ days. Random Forest (RF) algorithms
are used to perform TMR on a minute-by-minute basis. This study focuses on the
participant-level split and the proposed method demonstrates varying prediction
rates, achieving a prediction rate of $95\%$ for biking, $65\%$ for public
transport, and an overall prediction rate of $90\%$. Notably, this study, like
the vast majority of hybrid studies, use early fusion to combine these sensor
modalities by either extracting statistical features or concatenating the raw
signals.

To date, most approaches use location and inertial sensor data collected at high
sampling rates; in most studies, location sampling rate falls within the range
of $1$ to $5$ seconds \cite{Muhammad2021, Sadeghian2021}, while the sampling
rate of inertial data ranges between $20$ and $30$ Hz
\cite{Straczkiewicz2021}. Such requirements tend to have a strong impact on
modern smartphones energy consumption. Some studies have suggested that inertial
sensors sampled at $20$ Hz can sufficiently discriminate between various TMs
\cite{Gjoreski2020} and a sampling rate of $10$ Hz is adequate for
distinguishing between different mobility patterns \cite{
  Wannenburg2016}. Notably, one study demonstrated that reducing the sampling
rate from $100$ Hz to $12.5$ Hz led to a threefold increase in the duration of
data collection on a single battery charge \cite{Yurur2013}. Regarding location
data, power consumption decreases with increased sampling time and it is
estimated that employing a sampling time of $40$ seconds can, on average, extend
the mobile phone battery life by $25.4\%$ compared to the standard sampling
setting of $1$ Hz\cite{Castrogiovanni2020}.

Thus, one challenge in developing such a hybrid approach is to fuse the
different sensor modalities efficiently and maintain low sampling-rate
requirements. To address these challenges, we use accelerometer signals at $10$
Hz and location signals at $1/60$ Hz (one sample per minute) and propose a
hybrid attention-based MIL system that extracts instance-level embeddings
through single-sensor--based feature encoders and fuses the instance-level
features into joint representations through a deep MIL mechanism
\cite{Ilse2018}. The network's ability of being selective enables it to
effectively fuse the acceleration and location modalities, adeptly addressing
situations where location signal is not available by solely relying on
accelerometer data. Morevoer, the network's interpretability capabilities offer
a rich spectrum of insights into the TMR task. The classification task includes
eight TMs: still, walk, run, bike, car, bus, train, and subway, and the model is
evaluated with LOSO CV, achieving accuracy and F1-Score of $90.8\%$ and
$92.6\%$, respectively.

\section{Materials and methods}
\label{sec:materials}

Our approach is based on an artificial neural network (ANN) that separately
processes windows of acceleration and location signal and then combines them
with MIL. Each signal type is originally pre-processed and then windows are
extracted. We use two different sub-networks for extracting features from each
signal type and then map them to a common embedding feature space. A third
sub-network is then used that performs attention-based MIL. Finally, we
post-process the predicted labels using an HMM.

\subsection{Acceleration pre-processing}
\label{subsec:acceleration-pre-processing}

While it is possible to use high sampling rates to capture acceleration in
modern devices, continuous capturing at such rates can have a detrimental effect
on battery life. Thus, we opt to sample acceleration at a low rate of
$f_s^a = 10$ Hz. We extract consecutive windows of $T^{a}=60$ sec. This choice
is based on the assumption that TM does not change significantly within a single
minute. Experimenting with the dataset also shows similar results (see Section
\ref{subsubsec:pre-processing}).

The accelerometer provides 3D measurements:
$\mathbf{a}[n] = [a_x[n], a_y[n], a_z[n]]^{T}$. To remove the effect of phone
and sensor orientation, we compute the magnitude
$a[n] = \lVert \mathbf{a}[n] \rVert$ \cite{Ahmed2019}. We also compute the
magnitude of jerk as:
\begin{equation}
  j[n] = \lVert \mathbf{a}[n] - \mathbf{a}[n-1] \rVert \cdot f_{s}^a
\end{equation}
because it is orientation-independent and reflects body-related accelerations
\cite{Wilhelmiina2011}. Thus, we obtain a signal of $2$ channels:
$[a[n], j[n]]^{T}$.

It is important to note that we do not remove the gravity component from the
mangitude $a[n]$, as it proves to be a source of valuable information (based on
early-stage experimentations with the dataset). However, the effect of gravity
is indirectly removed when computing jerk, $j[n]$. Thus, our method takes
advantage of both versions of the signal.

Finally, we transform each of the $2$ channels into spectrograms, using the
short-time Fourier transform (STFT). Power spectrum is estimated in $10$-second
segments with $9$ sec overlap, yielding $51$ segments in total. We
rescale the power spectrum bands into $51$ bands by grouping frequencies; each band has
double bandwidth compared to the previous one, i.e.:
\begin{equation}
  \frac{f_{b}[i + 1] - f_{b}[i]}{f_{b}[i]-f_{b}[i-1]}=2,\, i=1, \ldots, 50
\end{equation}
where $f_{b}[i-1]$ and $f_{b}[i]$ are the lowest and highest frequency of the
$i$-th band. The spectrograms are then concatenated as two channels, resulting
in a single ``image'' of $51 \times 51 \times 2$.

To reduce the risk of overfitting, particularly in cases when there is a
considerable overlap between training samples, and to enhance the generalization
capabilities of our ANN, we incorporate data augmentation.  Typically,
accelerometer signals are augmented using signal processing transformations such
as jittering, scaling, rotation, and permutation \cite{Um2017,
  Kalouris2019}. While these methods have shown improvements in recognition
rates, they tend to disregard the correlations between signals, limiting the
potential to uncover cross-signal features. We adopt an alternative augmentation
policy, proposed by Park \cite{Park2019}. Inspired by cut-out
\cite{DeVries2017}, consecutive time steps or frequency channels within the
spectrogram are randomly masked, reducing reliance on specific regions or
features while preserving the intrinsic cross-signal correlations. In
particular, we randomly choose $0$, $1$, or $2$ frequency bands of random width
($0$ to $5$ rows), and also randomly choose $0$, $1$, or $2$ time intervals of
random length ($0$ to $5$ columns),and mask them.

\subsection{Location pre-processing}
\label{sec:location_preprocessing}

Most studies use location data captured at high sampling rates (e.g. $1$
Hz). Such rates have a notable impact on battery usage, even in modern
smartphones. Thus, we opt for a reduced sampling rate of $f_s^l = 1/60$ Hz,
which translates to one sample per minute, in order to simulate a more
battery-friendly data collection approach.

Capturing location data depends on GPS signal availability, resulting in certain
sections of the time-series data being frequently absent due to poor satellite
reception. We identify two types of ``gaps'': (a) brief signal losses, lasting
less than $3$ minutes, for which we employ linear interpolation to estimate the
missing values, and (b) extended periods of signal loss, exceeding $3$ minutes,
during which we apply a masking value the missing values that signifies the
presence of missing data. This approach is necessary since linear interpolation
would yield erroneous data in such cases.

The signal is then segmented to windows, similarly to the accelerometer
signal. However, we use windows of $12$ minutes. This value of $12$ minutes
enables us to capture behavior that is not observable in shorter durations
(i.e., moving through heavy traffic in a motor vehicle can create patters that
last longer than $1$ minute). It should be also noted that choosing smaller
window sizes (such as $1$ minute as in the case of the acceleration signal)
would yield windows with very few signal samples (e.e., only one for 
$1$-minute windows).

The big difference in window lengths between the two modalities as well as the
orders of magnitude of difference between the sampling rates creates a challenge
when trying to combine tehm in a single detector. We face this challenge by
designing our architecture appropriately (please see Section
\ref{subsec:embedding} for more details).

Using raw location coordinates as input data is not practical as it may lead to
the development of a location-dependent classifier. To overcome this we
transform the coordinates into location-invariant features. Specifically, for a
window of location signal, $l[n] = (\text{lat}[n], \text{lon}[n])$, at
timestamps $t[n]$ for $n=0, \ldots, 11$, we consider $2$ features, namely speed as:
\begin{equation}
  v^{l}[n] = \frac{d(l[n], l[n - 1])}{t[n] - t[n - 1]}
\end{equation}
and acceleration as:
\begin{equation}
  a^{l}[n] = \frac{v^{l}[n] - v^{l}[n - 1]}{{t}[n] - {t}[i - 1]}
\end{equation}
where $d(\bullet,\bullet)$ is the Havershine distance between two pairs of
coordinates. Because of how $v^{l}$ and $a^{l}$ are computed, the result is a
$10 \times 2$ matrix. Additionaly, we calculate $5$ more features for each window:
mean and standard deviation of speed, mean and standard deviation of speed
derivative, and ``movability'' \cite{ijgi6030063}. Movability indicates the
ratio of total displacement to the total distance covered within the examined
period:
\begin{equation}
  m = \dfrac{d(l[11], l[0])}{\sum\limits_{k=1}^{k=11}d(l[k], l[k-1])}
\end{equation}

\subsection{Embedding signals into a common space}
\label{subsec:embedding}

To effectively integrate the information from the two sensor modalities into a
cohesive joint representation, we must account for the variations between these
modal inputs, including their distinct nature, representation, sampling rate,
and window size. To tackle these challenges we adopt a two-pronged approach,
employing two parallel modality-specific branches to embed the information
originating from acceleration and location into a common, embedding space. To
achieve this, we project the data of Sections
\ref{subsec:acceleration-pre-processing} and \ref{sec:location_preprocessing}
onto $R^d$ (where $d = 256$ in our case), using two feature encoders:
\begin{itemize}
\item
  $\mathbf{f}_{a}: \mathbb{R}^{51 \times 51 \times 2} \rightarrow \mathbb{R}^{d}$, the acceleration-feature encoder
\item $\mathbf{f}_{l}: \mathbb{R}^{10 \times 2} \rightarrow \mathbb{R}^{d}$, the location-feature encoder
\end{itemize}
It is important to note that we have designed the two modality-specific networks
so that their output is the same, i.e., both produce vectors in $\mathbb{R}^{d}$ where
$d=256$. Thus, windows of both signal types are embedded in the same feature
space, enabling the application of MIL.

\begin{figure}
  \centering
  \includegraphics[width=.95\linewidth]{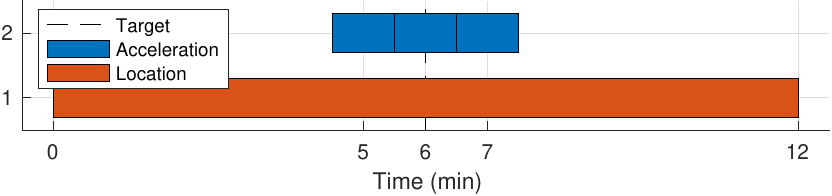}
  \caption{Visual representation of the windows that are used for a single bag
    (for MIL). Given a target timestamp (dashed line), we include $n_l = 1$
    window of location data that is $12$ minutes long and $n_a = 3$ successive
    windows of acceleration data that are each $1$ minute long. The next bag is
    obtained by shifting everything by $1$ minute.}
  \label{fig:windows-plot}
\end{figure}

Thus, we create bags of $n = 4$ instances, including $n_a = 3$ acceleration
windows and $n_l = 1$ location windows, as shown in Figure
\ref{fig:windows-plot}. This, combined with MIL, enables us to fuse the two
types of signals despite their very different nature.

The acceleration-feature encoder, $\mathbf{f}_{a}$, is a convolution model. It
consists of an initial batch-normalization layer, followed by $3$ convolutional
blocks, and then $2$ fully-connected (FC) blocks. Each convolutional block
consists of a 2D-convolutional layer with $3\times3$ kernels, batch normalization,
Re.L.U. activation, and finally max-pooling (ratio $2:1$ per dimension and
stride of $2$). The $3$ convolutional blocks have $16$, $32$, and $64$ filters
respectively. A FC block consists of a dropout layer, a FC layer, batch
normalization, and Re.L.U. activation. The first block acts as a bottleneck, by
decreasing the dimension to $128$. The second block increases the dimension to
$d$.


The location-feature encoder, $\mathbf{f}_{l}$, combines a bi-directional LSTM
of $128$ cells with the set of hand-crafted features (Section
\ref{sec:location_preprocessing}) with $3$ FC blocks. Prior to the Bi-LSTM
layer, batch normalization is applied. The output of the Bi-LSTM layer, which
includes both the first and last states of the Bi-LSTM, is concatenated with $5$
hand-crafted features that are computed during the feature extraction
phase. Three fully-connected blocks follow; each block consists of a
fully-connected layer, batch normalization, and then Re.L.U. activation.


\subsection{Multiple instance learning and classification}
\label{subsec:MIL}

Our main focus is leveraging MIL to effectively integrate information from the
two types of modalities. We use bags of $n$ instances, including $n_a$
acceleration instances and $n_l$ location instances, as shown in Figure
\ref{fig:windows-plot}. However, directly combining acceleration and location
instances in their raw form is not a feasible solution due to their inherent
differences, such as different representation and captured information. To
address this challenge, we map them into a common embedding space, as discussed
in Section \ref{subsec:embedding}. The acceleration feature encoder
$\mathbf{f}_a$ transforms the acceleration instances
$S_a = \lbrace \mathbf{s}_{a,1}, \dotsm, \mathbf{s}_{a,n_a} \rbrace \in \mathbb{R} ^ {n_a \times 51 \times 51 \times
  2}$ into a set of acceleration-based embeddings
$H_a = \lbrace \mathbf{h}_{a,1}, \dotsm, \mathbf{h}_{a,n_a} \rbrace \in \mathbb{R} ^ {n_a \times
  d}$. Similarly, the location feature encoder transforms the location instances
$S_l = \lbrace \mathbf{s}_{l,1}, \dotsm, \mathbf{s}_{l,n_l} \rbrace \in \mathbb{R} ^ {n_l \times 10 \times 2}$
into a set of location-based embeddings
$H_l = \lbrace \mathbf{h}_{l,1}, \dotsm, \mathbf{h}_{l,n_l} \rbrace \in \mathbb{R} ^ {n_l \times d}$.

Subsequently, we introduce the MIL ANN
$\mathbf{f}: \mathbb{R}^{N \times d} \rightarrow \mathbb{R}^{d}$, which aggregates the multi-modal instance-level
embeddings
$H = \lbrace \mathbf{h}_{a,1}, \dotsm, \mathbf{h}_{a,n_a}, \mathbf{h}_{l,1}, \dotsm,
\mathbf{h}_{l,n_l} \rbrace \in \mathbb{R} ^ {N \times d}$ into a fused encoding of the same dimension
$d = 256$.


To emphasize the expressiveness of each instance and prioritize the most useful
ones, we propose a weighted average of the instance-level embeddings. Inspired
by the work of Ilse \cite{Ilse2018} as well as of Papadopoulos
\cite{papadopoulos2020detecting}, we design an attention-based fusion network
responsible for determining the instance weights $a_n$ for the embedded
instances $\mathbf{h}_{n}$. This network can be formalized as follows:
\begin{equation}
  a_{n} = \frac
  {e^{\left( \mathbf{w}^{T} \tanh{(\mathbf{V} \mathbf{h}_{n}^{T})} \odot \sigm{(\mathbf{U} \mathbf{h}_{n}^{T}
          )} \right)}}
  {\sum_{j=1}^{N} e^{\left( \mathbf{w}^{T}\tanh{(\mathbf{V} \mathbf{h}_{j}^{T})} \odot \sigm{(
          \mathbf{U} \mathbf{h}_{j}^{T})} \right)}}
\end{equation}
where $\mathbf{w} \in \mathbb{R} ^ {256 \times 1}$,
$\mathbf{V} \in \mathbb{R} ^ {256 \times d}$, and
$\mathbf{U} \in \mathbb{R} ^ {256 \times d}$ are trainable parameters of the attention network,
and $\odot$ is the Hadamard product operator. Accordingly, the contribution of the
individual instances is subject to the  attention weights
$\lbrace a_{a,1}, \dotsm, a_{a,n_a}, a_{l,1}, \dotsm, a_{l, n_l} \rbrace$. By definition, it
always holds that $\sum_{i=1}^{n_a}a_{a,i} + \sum_{i=1}^{n_l}a_{l,i} = 1$, where the
first term indicates the importance attributed to the acceleration modality,
while the second term represents the importance of the location
modality. Consequently, a modal-fused encoding $\mathbf{z}$ is derived by
calculating the weighted sum of the instance-level embeddings as:
\begin{equation}
  \mathbf{z} = \sum_{n=1}^{N} a_{n} \cdot \mathbf{h}_{n} =
  \sum_{i=1}^{n_a} a_{a,i} \cdot \mathbf{h}_{a,i} +  \sum_{i=1}^{n_l} a_{l,i} \cdot \mathbf{h}_{l,i}
\end{equation}

\begin{figure*}
  \centering
  \includegraphics[width=.8\linewidth]{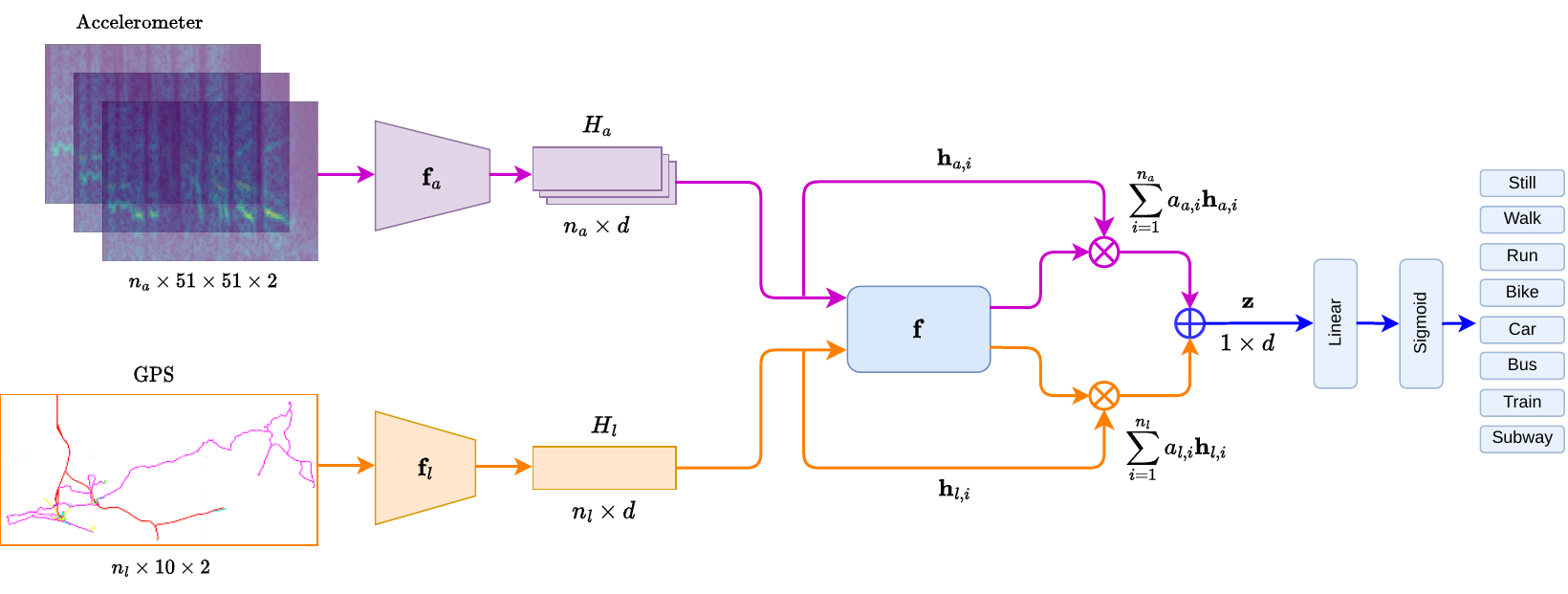}
  \caption{Proposed multi-modal TMR network. The input is a bag containing two
    types of instances: acceleration ($n_a$) and location ($n_l$). Instances are
    processed in parallel by two modality-specific feature encoders
    $\mathbf{f}_a$ and $\mathbf{f}_l$ accordingly, which embed them in the same
    $d$-dimensional space. The attention-based MIL ANN $\mathbf{f}$, follows; it
    aggregates $n$ instances (of the $\mathbb{R}^{d}$ embedding space) into a fused
    attentive encoding of the same dimension $d$. Finally, to make
    transportation mode predictions, a small classification network maps the
    fused encoding $\mathbf{z}$ into category-wise decision scores. }
  \label{fig:architecture}
\end{figure*}

Finally, the classification ANN,
$\mathbf{c}: \mathbb{R}^{d} \rightarrow \mathbb{R}^{m}$, maps the fused attentive encoding
$\mathbf{z} \in \mathbb{R}^{d}$ to the label space (where $m=8$ in our case); It is a small
network, consisting of one fully-connected block (FC layer, batch normalization,
and Re.L.U. activation) followed by one FC layer. The problem at hand is
formulated as a multi-class classification task, in the sense that there is
always a single ``right'' mode to be recognized. However, given that certain
transportation modes, like ``train'' and ``walking'', are not mutually
exclusive, the sigmoid activation is employed to provide the final class
probability of the $j$-th TM $p_{j}$ (out of $8$ TMs).  A complete overview of
the entire architecture is shown in Figure \ref{fig:architecture}.

\subsection{Post-processing}
\label{sec:postprocessing}

While our ANN classifies each example (bag) independently, it is worth noting
that there are correlations in transportation mode patterns. For instance, it is
unlikely that we would immediately take the bus after riding the subway, without
walking in between. To account for such correlations and improve the accuracy of
the classification output, a common approach is to apply a smoothing operation,
such as a rolling median filter. In our work we employ an HMM that effectively
captures temporal dependencies inherent in transportation mode patterns to
determine the most probable sequence of classifications and to mitigate
classification noise. We apply the HMM in the following manner:
\begin{itemize}
\item States are defined as the TMs, i.e., a total of $8$ states are defined
\item State-transition probabilities $P(c_m|c_{m-1})$ are estimated from the
  transition matrix derived from the combined training and validation sets
\item Emission probabilities $P(c_m|\hat c_m)$ are estimated using the
  prediction probability estimates $p$
\item Start probabilities $P(c_1)$ are uniformly set to $1/8$ (since $8$ is the
  number of TM classes)
\item The HMM is applied to each recording session and the most likely sequence
  of predictions is computed
\end{itemize}

\section{Experimentation setup}
\label{sec:setup}

\subsection{Dataset}
\label{sec:dataset}

To evaluate our approach we employ the publicly available ``SHL Preview
Dataset''. This dataset is available as part of the complete Sussex-Huawei
Locomotion Dataset \cite{gjoreski2018university, Wang2019} and includes $59$
hours of annotated recordings; it was captured using $4$ different smartphones
(attached in $4$ different body positions) on $3$ participants, as they went
about their normal lives over the course of $3$ days. This dataset contain $8$
TMs: still, walk, run, bike, car, bus, train, and subway, and preserves all
sensor modalities except for audio. In our study, we only consider the
accelerometer and location sensor modalities. Acceleration signals are
downsampled from the original $100$ Hz to $10$ Hz and location signals are
downsampled from $1$ Hz to $1/60$ Hz. The processed acceleration signals are
segmented into discrete frames, each with $600$ samples ($60$ sec). At this
sample rate, the dataset yields a total of $3,421$ acceleration frames,
resulting in a total data size of $3,421 \times 600$ for each sensor placement. The
transportation mode labels are synchronized with the acceleration frames,
yielding a total of $3,421$ labels. Location signals are segmented into
$12$-minute windows, each containing $12$ samples with an $11$-minute overlap
with neighboring windows. Location data is recorded asynchronously with average
availability of location signal of $80\%$. In total, $2,715$ location samples
are obtained against the overall $3,421$ acceleration frames and labels.

In our scenario, we divide the dataset using LOSO; each user's data is used in
turn for testing and the data from the remaining users are used for model
development. The development data (for each LOSO iteration) consist of data from
$2$ participants. We split into $80\%$ training set and $20\%$ validation set
using a stratified split framework \cite{Kenneth1999} so that both sets maintain
the same label and user distribution.

However, random splitting may produce heavily dependent sets since pairs of
highly correlated neighboring windows could be assigned to the training and
validation set respectively \cite{Widhalm2018}. To avoid this, we select long,
continuous streams of data (instead of individual windows) as the unit for
random splitting. It is important to note that, since both training and
validation data are captured from the same subjects, effectiveness on the
validation set may be positively biased compared to the expected effectiveness
on the unseen test set.

\subsection{Model Training}
\label{sec:model-training}

While our proposed ANN (Figure \ref{fig:architecture}) can be trained
end-to-end, we opt to separately train the acceleration-feature encoder
$\mathbf{f}_{a}$. Subsequently, we freeze the trained weights and integrate them
in the complete ANN model and trainit, as described in Section
\ref{subsubsec:pre-training}. Both training steps are implemented in the same
manner. A standard categorical cross-entropy (CCE) is used as the loss function
to compute the error between the prediction probabilities and the true
labels. To minimize the CCE loss and update the trainable weights, the Adam
optimizer \cite{adam} is used with an initial learning rate of $10^{-4}$. We use
a batch size of $32$ and train for at most $80$ epochs. To avoid overfitting we
use early stopping to identify the ideal number of epochs for training the
network.

\section{Evaluation and Discussion}
\label{sec:results}

We evaluate effectiveness on the basis of accuracy and macro-averaged
F1-score. All subsequent experiments are conducted in a user-independent manner
(using LOSO). We also repeat each experiment $15$ times to account for random
initialization of model weights and random train and validation set split.

\subsection{Influence of device placement}
\label{subsec:placement}

To assess the effectiveness of our model and gain insights into the influence of
device placement, we propose a comprehensive assessment from three different
perspectives:
\begin{itemize}
\item{Per-placement: we exclusively train and test our model using accelerometer
    data obtained from a single smartphone position}
\item{All-placements: we train our model using data from all four distinct
    acceleration data streams, each corresponding to a different recording
    position. Subsequently, we evaluate the model's effectiveness using
    accelerometer data obtained exclusively from a single position}
\item{Mixed-placement: we generate new virtual streams by combining sequential
    acceleration data randomly obtained from different positions. This
    experiment aims to simulate real-world scenarios where the position of the
    smartphone changes over time. We conduct two experiments of this type; one
    utilizing a single mixed stream for training and another utilizing multiple
    mixed streams (four in total) for training. We include both experiments for
    a more thorough evaluation}
\end{itemize}

\begin{table}
  \caption{Results of the proposed approach on the eight-class task. Results are
    averages across $15$ runs.}
  \label{tab:results_8class}
  \centering
  \begin{tabular}{lcccc}
    \toprule
    \multirow{2}{*}{} & \multicolumn{2}{c}{\textbf{No HMM}}
    & \multicolumn{2}{c}{\textbf{HMM}}
    \\
    \cmidrule{2-3} \cmidrule{4-5} & \textbf{Accuracy} & \textbf{F1-score} & \textbf{Accuracy}
                                                      & \textbf{F1-score}
    \\
    \cmidrule{1-5}
    \multicolumn{5}{l}{\textit{Per-placement}} \\
    \cmidrule{1-5}
    Bag      & $88.0$ & $84.9$ & $91.5$ & $87.3$ \\
    Hand     & $82.0$ & $81.3$ & $87.7$ & $85.6$ \\
    Hips     & $85.5$ & $80.3$ & $88.7$ & $82.0$ \\
    Torso    & $84.9$ & $79.4$ & $88.4$ & $83.3$ \\
    Average  & $85.1$ & $81.5$ & $89.1$ & $84.6$ \\
    \cmidrule{1-5}
    \multicolumn{5}{l}{\textit{All-placements}} \\
    \cmidrule{1-5}
    Bag      & $90.5$ & $89.0$ & $94.6$ & $92.6$ \\
    Hand     & $84.8$ & $84.5$ & $90.3$ & $89.5$ \\
    Hips     & $89.1$ & $87.8$ & $93.0$ & $91.2$ \\
    Torso    & $88.2$ & $85.4$ & $92.5$ & $89.7$ \\
    Average  & $88.1$ & $86.7$ & $92.6$ & $90.8$ \\
    \cmidrule{1-5}
    \multicolumn{5}{l}{\textit{Mixed-placement}} \\
    \cmidrule{1-5}
    One      & $86.6$ & $84.8$ & $91.3$ & $88.6$ \\
    Multiple & $89.3$ & $87.9$ & $93.7$ & $92.3$ \\
    \bottomrule
  \end{tabular}
\end{table}

It is important to note that we do not include any information regarding the
placement of the acceleration sensor in any of the experiments mentioned
above. Instead, the model remains unbiased and agnostic towards sensor
placement, enabling independent learning and distinguishing of the sources of
the acceleration data. Furthermore, these experiments are specifically focused
on the accelerometer sensors. In contrast, location signals exhibit minimal to
no variation across different recording positions. This suggests that the
inclusion of location signal in the recognition system enhances its robustness
to variations in position.

Based on Table \ref{tab:results_8class} there is a clear ranking of sensor
placement: Bag $>$ Hips $>$ Torso $>$ Hand. The data collected from the Hand
placement appears to be noisier compared to the other three placements. This is
likely due to the physical contact and interaction between the hand and the
phone, as it introduces more variability and interference in the captured
data. Additionally, the Hand placement seems to be more sensitive to individual
user variations and the way each subject uses their smartphone. On the contrary,
placing the phone in a Bag yields higher inference accuracy since this placement
involves less movement variability and is in close proximity to the body. Hips
and Torso exhibit similar behavior.

Table \ref{tab:results_8class} also demonstrates a significant improvement in
effectiveness for the generalized classifier when compared to the per-placement
experiment. Evidently, in the second experiment, the model is trained across
different placements, thus learning more general features, while reducing the
risk of overfitting the data from a specific placement.

One might expect the effectiveness in the mixed-placement experiment to be on
par with the average effectiveness of the all-placements experiment, since both
experiments utilize all the available data from every recording
placement. However, our model can benefit from the mixed streams and the dynamic
smartphone position; the attention-based MIL mechanism can effectively allocate
its attention among a sequence of acceleration windows obtained from different
positions and leverage the changing position in its advantage (see Section
\ref{subsec:model-interpretability}). For instance, if a sequence contains two
windows obtained from the Hand position and one obtained from the Bag position,
the model will likely prioritize the window acquired from the Bag position. This
increase in effectiveness is evident on Table \ref{tab:results_8class}.


\subsection{Acceleration-specific study}
\label{subsec:unimodal}

In all subsequent studies, we maintain the structure of the second experiment
(as mentioned in \ref{subsec:placement}), and the effectiveness is reported as
the average score across all testing positions. To gain insights into the
acceleration-specific aspect of our proposed model, we examine the impact of
three important components: the pre-processing pipeline, using MIL to exploit
acceleration data, and the importance of data augmentation.

\subsubsection{Pre-processing pipeline}
\label{subsubsec:pre-processing}

Initially, we conduct tests using various sampling frequencies and the findings
indicate that the highest  accuracy is attained at a sampling rate of $10$
Hz. We, then, explore the importance of the acceleration-derived signals,
comparing all potential axes $\mathbf{a}[n] = [a_x[n], a_y[n], a_z[n]]^{T}$,
along with the magnitude $a[n]$ and the jerk norm $j[n]$. Individually, the
magnitude outperforms all other signals, while the jerk exhibits the poorest
effectiveness. However, when combined, the magnitude and jerk yield better
results compared to using the magnitude alone, and slightly worst to utilizing
all the signals together.  Next, we compare the different parameters involved in
generating the 2D spectrogram representations. We observe that employing the
logarithm of the power yields superior results compared to using the raw
power. Additionally, a logarithmic interpolation for the frequency axis proves
to be an effective approach for reducing data size while retaining essential
information. Finally, after experimenting with different short-time Fourier
transform (STFT) window durations, we conclude that the highest effectiveness is
achieved for a window with a duration of $10$ seconds.

\subsubsection{MIL for acceleration}
\label{subsubsec:acceleration-MIL}

In this study, we employ MIL to leverage sequences of acceleration windows,
instead of relying solely on individual acceleration windows.


To assess the importance of this method, we compare three variations of our
model:
\begin{itemize}
\item Acc-CNN: A single $d$-minute acceleration window is fed
  into Acc-CNN. This network consists of the acceleration-feature encoder
  $\mathbf{f}_{a}$ (see Section \ref{subsec:embedding}) followed by a
  classification layer (FC layer and sigmoid activation) that maps the final
  embedding $\mathbf{h}_{a}$ into class probabilities $p$.
\item Acc-MIL: The same $d$-minute window is now segmented into a sequence of
  $d$ successive one-minute windows, which are fed to Acc-MIL. Similar to
  Acc-CNN, Acc-MIL also includes the acceleration-feature encoder
  $\mathbf{f}_{a}$ but additionally incorporates the MIL ANN $\mathbf{f}$,
  followed by the same classification layer.
\item Fusion-MIL: A sequence of $d$ one-minute acceleration windows along with a
  single $12$-minute location window, as depicted in Figure
  \ref{fig:windows-plot}. The proposed Fusion-MIL model exploits this bag of
  multi-modal instances to perform TMR, as illustrated in Figure
  \ref{fig:architecture}
\end{itemize}

\begin{figure}
  \centering
  \includegraphics[scale=0.4]{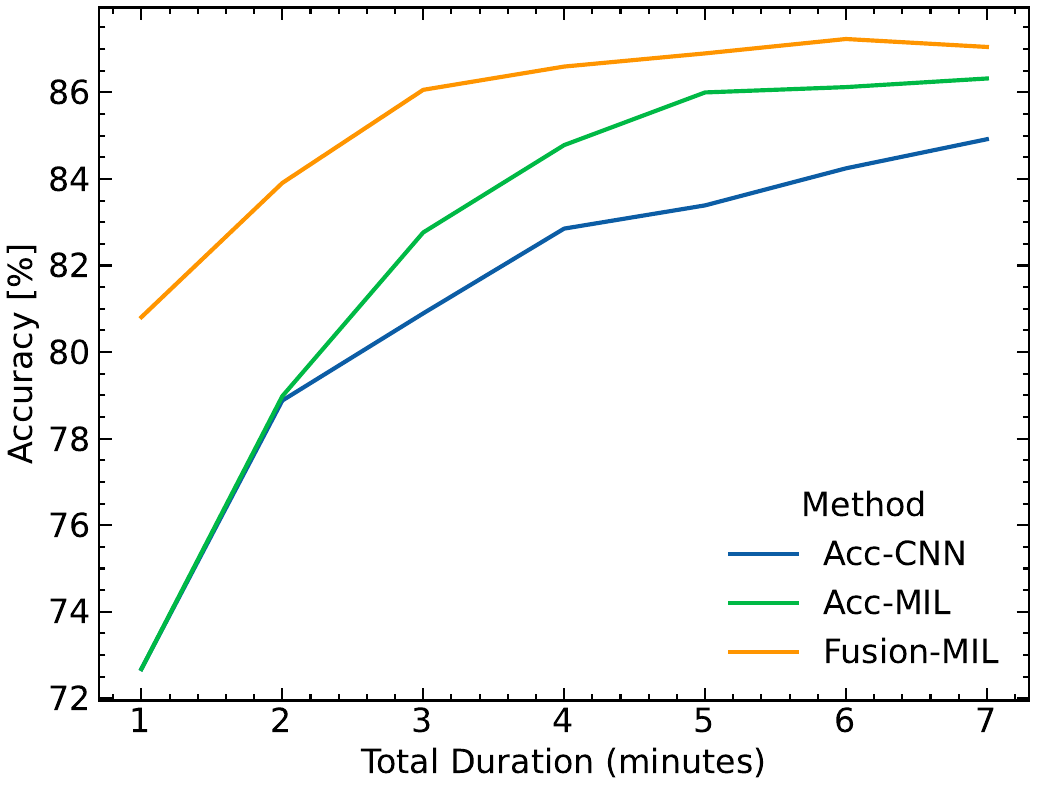}
  \caption{Accuracy comparison between Acc-CNN, Acc-MIL and Fusion-MIL versus
    the duration $d$ of acceleration input data}
  \label{fig:ACC-MIL_acc}  
\end{figure}

In Figure \ref{fig:ACC-MIL_acc}, we present a effectiveness comparison between
these three approaches as the duration $d$ of the input increases. Evidently,
removing MIL causes a clear effectiveness drop.

\subsubsection{Data augmentation}

\begin{figure}
  \centering
  \label{fig:aug_acc}
  \includegraphics[scale=0.4]{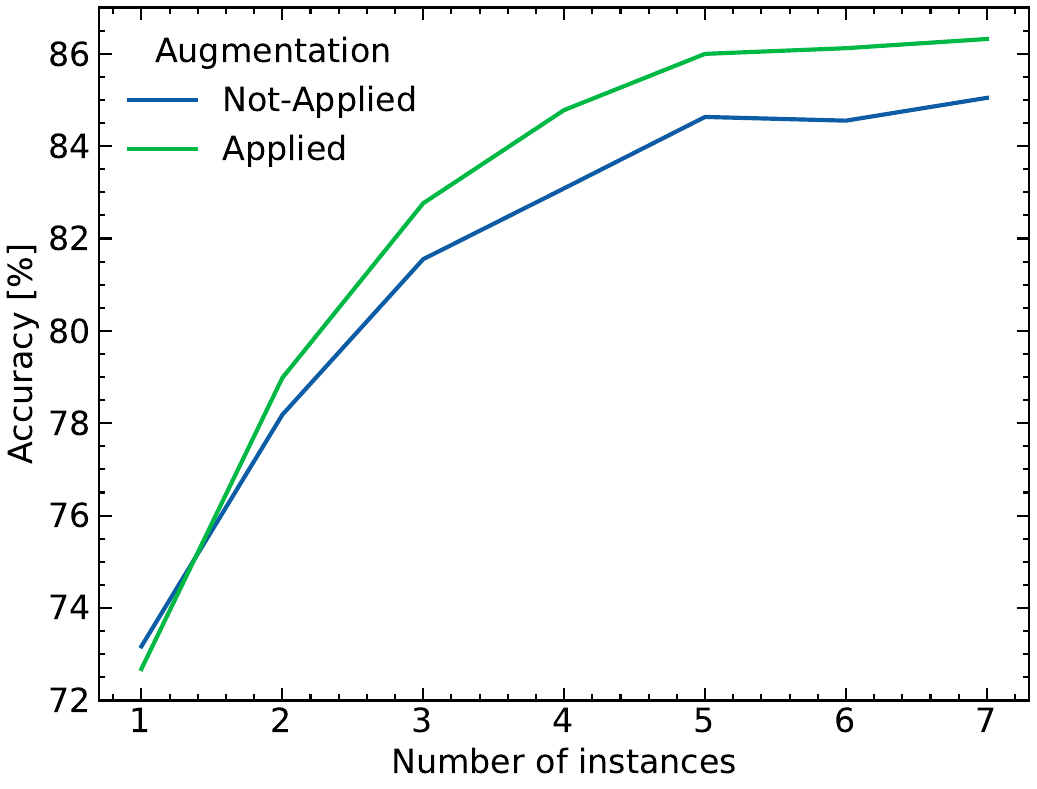}
  \caption{Effect of data augmentation on the accuracy of Acc-MIL versus the
    number $d$ of one-minute acceleration instances}
\end{figure}

As we progressively increase the number of acceleration instances, we
inadvertently increase the overlap across different examples, increasing the
risk of over-fitting. To mitigate this issue and improve generalization, we opt
to augment the acceleration data before feeding it into the model (see
section~\ref{subsec:acceleration-pre-processing}). Figure \ref{fig:aug_acc}
investigates the influence of data augmentation on the effectiveness of Acc-MIL
as the number of acceleration instances increases.

\subsection{Multi-modal study}
\label{subsec:multimodal}

We assess the effect of three significant designs, i.e. alternative
approaches to Fusion-MIL, pre-training, and Markov smoothing.

\subsubsection{Comparison with alternative approaches}
\label{subsubsec:baselines}

We compare the classification effectiveness of our proposed Fusion-MIL model to
the following single-modal and multi-modal approaches (results in Table
\ref{tab:baselines}):
\begin{itemize}
\item Loc-LSTM: A location-only ANN with a $12$-minute location window as its
  input. This network consists of the location-feature encoder $\mathbf{f}_{l}$
  followed by a classification layer (FC layer and sigmoid activation) that maps
  the final embedding $\mathbf{h }_{l}$ into class probabilities $p$.
\item Acc-CNN: A acceleration-only ANN with a $3$-minute acceleration window as
  its input (see Section \ref{subsubsec:acceleration-MIL}).
\item Fusion-Concat: A multi-modal network that accepts a $3$-minute
  acceleration window and a $12$-minute location window as a paired input. It
  combines the acceleration-feature encoder $\mathbf{f}_{a}$ and the
  location-feature encoder $\mathbf{f}_{l}$, followed by a fusion layer that
  concatenates the $256$-dimensional modal embeddings $\mathbf{h}_{a}$ and
  $\mathbf{h}_{l}$. The modal-fused $512$-dimensional encoding $\mathbf{z}$, is
  fed to the classification ANN $\mathbf{c}$ to derive the final classification
  probabilities.
\item Acc-MIL: A acceleration-only ANN that uses MIL to leverage a bag of
  $n_a = 3$ successive one-minute acceleration windows for classification (see
  ~\ref{subsubsec:acceleration-MIL}).
\item Fusion-Concat++: We enhance Fusion-Concat, by substituting its original
  $\mathbf{f}_{a}$ sub-network with Acc-MIL. The Acc-MIL branch exploits a set
  of $n_a = 3$ successive one-minute acceleration windows, mapping them into a
  joint $256$-dimensional embedding $\mathbf{h}_{a}$, while the $\mathbf{f}_{l}$
  branch maps a $12$-minute location window into a $256$-dimensional embedding
  $\mathbf{h}_{l}$. The $2$ embeddings are fused through the concatenation layer
  and the resulting modal-fused encoding $\mathbf{z}$ is mapped to the label
  space.
\end{itemize}

\begin{table}
  \centering
  \caption{Effectiveness comparison with alternative single-modal (SM) and
    multi-modal (MM) methods}
  \label{tab:baselines}
  \begin{tabular}{l@{\qquad}cc}
    \toprule
    Method                       & \textbf{Accuracy} & \textbf{F1-score} \\
    \midrule
    \textit{without MIL} \\
    \midrule
    Loc-LSTM (\textit{SM})        & $72.3$            & $61.4$            \\
    Acc-CNN (\textit{SM})         & $81.2$            & $80.7$            \\
    Fusion-Concat (\textit{MM})   & $84.5$            & $81.3$            \\
    \midrule
    \textit{with MIL} \\
    \midrule
    Acc-MIL (\textit{SM})         & $82.5$            & $81.9$            \\
    Fusion-Concat++ (\textit{MM}) & $85.2$            & $82.5$            \\
    Fusion-MIL (\textit{MM})      & $\mathbf{86.1}$   & $\mathbf{84.1}$   \\
    \bottomrule
  \end{tabular}
\end{table}

\subsubsection{Pre-training}
\label{subsubsec:pre-training}

Training the proposed ANN end-to-end reduces effectiveness, primarily due to the
limited availability of paired acceleration-location data in comparison to the
available acceleration-only data (the same location recording corresponds to $4$
different accelerometer recordings from $4$ different smartphone
positions). Thus, we opt to separately pre-train the uni-modal feature encoders
and then use the trained weights in the full ANN model:
\begin{itemize}
\item Acceleration encoder pre-training: To train $\mathbf{f}_{a}$ we adopt the
  Acc-CNN framework (Section \ref{subsubsec:acceleration-MIL}) with $1$-minute
  acceleration windows. After training $\mathbf{f}_{a}$ with all available
  training acceleration data we discard the classification layer and train the
  multi-modal ANN. We randomly use only one acceleration-location pair (out of
  $4$) per instance. To prevent over-fitting, we freeze the trained weights of
  $\mathbf{f}_{a}$ and solely train the remaining layers. This approach uses all
  available data and reduces the risk of overfitting.
\item Location encoder pre-training: We pre-train $\mathbf{f}_{a}$ following a
  similar process; we include this baseline for a more comprehensive evaluation.
\item Both encoders pre-training: We pre-train both acceleration and location
  feature encoders.
\end{itemize}

Table \ref{tab:pre-training} provides a comparison of incorporating pre-trained
uni-modal encoders. The results clearly indicate that the acceleration
pre-trained model surpasses the no pre-training model. By pre-training both
encoders there is only a marginal effectiveness improvement compared to solely
pre-training the acceleration encoder.

\begin{table}
  \centering
  \caption{Effect of pre-training}
  \label{tab:pre-training}
  \begin{tabular}{l@{\qquad}cc@{\quad}}
    \toprule
    Encoder pre-training                               & \textbf{Accuracy} & \textbf{F1-score} \\
    \midrule
    No pre-training                                    & 86.1              & 84.1              \\
    Location encoder $\mathbf{f}_{l}$                  & 87.0              & 84.8              \\
    Acceleration encoder $\mathbf{f}_{a}$              & 88.1              & 86.7              \\
    Both encoders $\mathbf{f}_{a}$ \& $\mathbf{f}_{l}$ & \textbf{88.8}     & \textbf{87.4}     \\
    \bottomrule
  \end{tabular}
\end{table}

\subsubsection{Smoothing with HMMs}
\label{sec:HMM}

\begin{figure}
  \centering
  \includegraphics[scale=0.55]{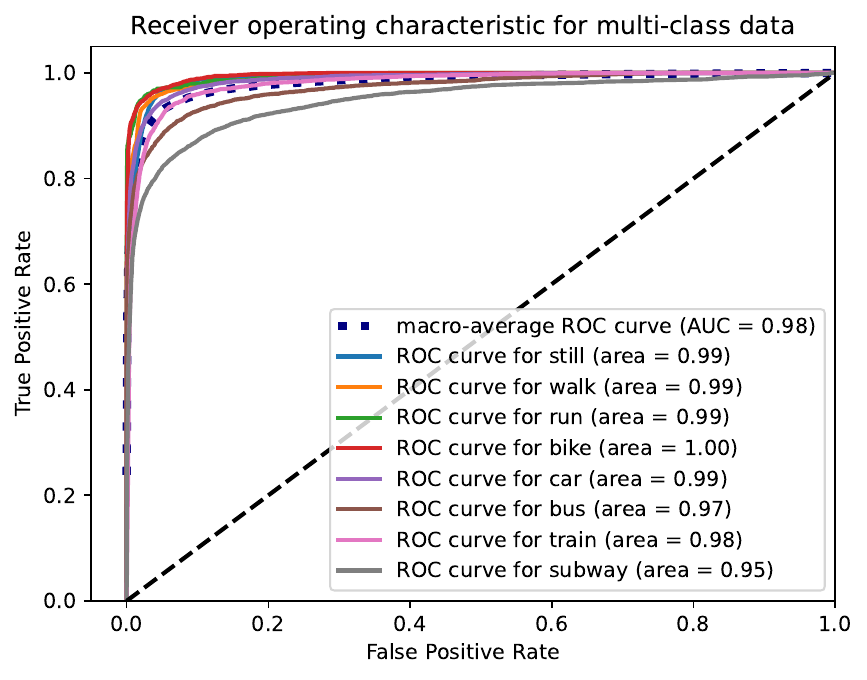}
  \caption{Multi-class ROC curves for Fusion-MIL}
  \label{fig:ROC_AUC}
\end{figure}

To determine the most probable sequence of TMs we employ Markov smoothing, which
reduces classification noise by exploiting the temporal correlation between
neighbouring samples. The HMM uses the prediction probability estimates ($p$) as
emission probabilities $P(c_m|\hat c_m)$ while the transition probabilities
$P (c_m|c_{m - 1})$ are estimated using the combined training and validation
sets. Table \ref{tab:post-processing} presents the effect of applying HMM
smoothing and Figure \ref{fig:ROC_AUC} shows the ROC curve without the HMM. It
can be observed that the area under the curve is highest for non-motorized TMs
(still, walk, run and bike) and slightly lower for motorized modes of
transportation.

\begin{table}
  \centering
  \caption{Overall effectiveness of our proposed method. Averages and standard
    deviations are computed over the $4$ placements and for each placement we
    perform $15$ runs with random initial model weights and train/validation
    split.}
  \label{tab:post-processing}
  \begin{tabular}{l@{\qquad}cccc@{\quad}}
    \toprule
    & \textbf{Accuracy} & \textbf{F1-score} & \textbf{Precision} & \textbf{Recall} \\
    \midrule
    no HMM     & $88.1 \pm 0.7$    & $86.7 \pm 1.0$    & $87.8 \pm 0.8$     & $86.6 \pm 1.2$  \\
    HMM        & $92.6 \pm 1.0$    & $90.8 \pm 1.5$    & $92.8 \pm 1.0$     & $90.6 \pm 1.6$  \\
    \bottomrule
  \end{tabular}
\end{table}

\subsection{Model interpretability}
\label{subsec:model-interpretability}

The significance of individual instances is determined by their respective
attention weights, i.e.:
$\lbrace a_{a,1}, \dotsm, a_{a,n_a}, a_{l,1}, \dotsm, a_{l, n_l} \rbrace$. These attention
weights play a crucial role in determining the contribution of each instance
within the model and enable the interpretability of our multi-modal model.

Using MIL on sequences of acceleration windows improves interpretability,
enhances the resolution of the acceleration input, and enables the
identification of the most relevant regions in the acceleration signal for the
final prediction.

Table \ref{tab:weight-std} presents the average standard deviation of attention
weights $\lbrace a_{a,1}, \dotsm, a_{a,n_a} \rbrace$ per class. This analysis enables us to
assess whether the MIL ANN evenly allocates its attention across all
acceleration instances within a bag or not. For periodic activities, such as
``walking'' or ``running'', we anticipate the generation of repetitive
acceleration recordings and similar instance-level embeddings
$\lbrace \mathbf{h}_{a,1}, \dotsm, \mathbf{h}_{a,n_a} \rbrace$. Accordingly, we would expect
uniformly distributed attention weights
$\lbrace a_{a,1}, \dotsm, a_{a,n_a} \rbrace$, resulting in low standard deviation, as
verified in Table \ref{tab:weight-std}. On the contrary, transportation modes
like ``car'' or ``bus'' are expected to produce dissimilar instances (such as a
``stopped'' instance followed by one ``moving'' instance), resulting in
diverging attention weights and high standard deviation, as verified in Table
\ref{tab:weight-std}. Basically, in such cases, sorting the acceleration
instances by their attention weights helps select the most ``active'' recordings
for further examination.

\begin{table}
  \centering
  \caption{Mean attention weight std. per acceleration bag}
  \label{tab:weight-std}
  \begin{tabular}{l@{\qquad}c@{\qquad}}
    \toprule
    \textbf{TM} & \textbf{Weight Std.} \\
    \midrule
    Still  & $0.041$              \\
    Walk   & $0.034$              \\
    Run    & $0.026$              \\
    Bike   & $0.042$              \\
    Car    & $0.058$              \\
    Bus    & $0.058$              \\
    Train  & $0.058$              \\
    Subway & $0.078$              \\
    \bottomrule
  \end{tabular}
\end{table}

Table \ref{tab:position-weights} presents the average position weights per
class, providing insight on how the MIL ANN allocates its attention among
acceleration instances in accordance to their recording positions. To obtain
these findings, we use the structure of the third experiment, which involves
mixed streams of acceleration recordings. This allows us to examine how the
model responds to bags composed of successive acceleration instances recorded in
various body positions.

\begin{table}
  \centering
  \caption{Mean attention weight per placement}
  \label{tab:position-weights}
  \begin{tabular}{l@{\qquad}cccc@{\qquad}}
    \toprule
    \textbf{TM}    & \textbf{Torso}   & \textbf{Hips} & \textbf{Bag}     & \textbf{Hand} \\
    \midrule
    Still          & $\mathbf{0.260}$ & $0.244$       & $0.253$          & $0.243$       \\
    Walk           & $\mathbf{0.278}$ & $0.236$       & $0.263$          & $0.223$       \\
    Run            & $\mathbf{0.256}$ & $0.250$       & $0.251$          & $0.242$       \\
    Bike           & $\mathbf{0.257}$ & $0.245$       & $\mathbf{0.257}$ & $0.241$       \\
    Car            & $0.256$          & $0.253$       & $\mathbf{0.262}$ & $0.229$       \\
    Bus            & $0.255$          & $0.253$       & $\mathbf{0.258}$ & $0.235$       \\
    Train          & $0.254$          & $0.252$       & $\mathbf{0.266}$ & $0.228$       \\
    Subway         & $0.260$          & $0.255$       & $\mathbf{0.262}$ & $0.223$       \\
    Average        & $\mathbf{0.259}$ & $0.249$       & $\mathbf{0.259}$ & $0.233$       \\
    \bottomrule
  \end{tabular}
\end{table}

As mentioned in Section \ref{subsec:MIL}, $\sum_{i=1}^{n_a}a_{a,i}$ indicates the
weight attributed to the acceleration modality, while $\sum_{i=1}^{n_l}a_{l,i}$
indicates the weight attributed to the location modality. Table
\ref{tab:modal-weights} presents the average modality weights per class to
provide insights on which modality the model relies the most when recognizing
particular TMs.

\begin{table}
  \centering
  \caption{Mean modality attention weight}
  \label{tab:modal-weights}
  \begin{tabular}{c@{\qquad}cc@{\qquad}}
    \toprule
    \multirow{2}{*}{\raisebox{-\heavyrulewidth}{Mode}} & \multicolumn{2}{c}{\textbf{Modality}} \\
    \cmidrule{2-3} & \textbf{Location} & \textbf{Acceleration} \\
    \midrule
    Still          & $0.355$           & $0.645$               \\
    Walk           & $0.384$           & $0.616$               \\
    Run            & $0.422$           & $0.578$               \\
    Bike           & $0.454$           & $0.546$               \\
    Car            & $0.461$           & $0.539$               \\
    Bus            & $0.426$           & $0.574$               \\
    Train          & $0.468$           & $0.532$               \\
    Subway         & $0.190$           & $0.810$               \\
    \bottomrule
  \end{tabular}
\end{table}

\subsection{Comparison against literature}
\label{subsec:comparison}

\begin{table*}[ht]
  \centering
  \caption{Summary of works using inertial and location data to detect
    transportation modes}
  \label{tab:sota_summary}
  \begin{tabular}{l|cccc|c|c|c|c|c}
    \toprule
    & \textbf{acc.} & \textbf{gyr.} & \textbf{mag.} & \textbf{loc.}
    & \textbf{Input}
    & \textbf{Sampling tate}
    & \textbf{Params ($10^{6}$)}
    & \textbf{Num. TMs}
    & \textbf{Reported F1-Score}
    \\
    \midrule
    CNN1D \cite{Liang2019} & \checkmark & & & & raw data & $50$ Hz & $0.3$ & $7$ & $94.5$ \\
    CNN2D \cite{Ito2018} & \checkmark & \checkmark & & & spectrogram & $100$ Hz & $0.5$ & $8$ & $88.8$ \\
    FPbiLSTM \cite{Tang2023} & \checkmark & \checkmark & \checkmark & & raw data & $20$ Hz & $3.1$ & $8$ & $94.2$ \\
    DT \cite{reddy2010} & \checkmark & & & \checkmark & features & $32$ Hz / $1$ sec & - & $5$ & $93.6$ \\
    Fusion-MIL (ours) & \checkmark & & & \checkmark & spectrogram + raw data & $10$ Hz / $60$ sec & $0.6$ & $8$ & $92.6$ \\
    \bottomrule
  \end{tabular}
\end{table*}

\begin{table}
  \caption{Comparison with literature}
  \label{tab:sota_comparison}
  \scalebox{0.88}{
    \begin{tabular}{lcccc}
      \toprule
      & \multicolumn{2}{c}{\textbf{No HMM}}
      & \multicolumn{2}{c}{\textbf{HMM}}
      \\
      \cmidrule{2-3} \cmidrule{4-5} & \textbf{Accuracy} & \textbf{F1-score} & \textbf{Accuracy}
                                                        & \textbf{F1-score}
      \\
      \cmidrule{1-5}
      \textit{Per-placement} \\
      \cmidrule{1-5}
      CNN1D \cite{Liang2019} & $55.3\pm10.3$ & $51.8\pm12.6$ & $63.3\pm10.6$ & $59.0\pm14.0$ \\
      CNN2D \cite{Ito2018} & $60.8\pm6.9$ & $59.5\pm6.8$ & $68.5\pm7.2$ & $65.0\pm7.2$ \\
      FPbiLSTM \cite{Tang2023} & $66.8\pm4.8$ & $65.0\pm4.9$ & $74.3\pm6.4$ & $70.0\pm5.8$ \\
      DT \cite{reddy2010} & $84.3\pm1.2$ & $72.5\pm2.8$ & $87.9\pm1.6$ & $79.6\pm2.5$ \\
      Fusion-MIL (ours) & $\mathbf{85}.1\pm2.7$ & $\mathbf{81.5}\pm2.4$ & $\mathbf{89.1}\pm1.7$ & $\mathbf{84.6}\pm2.4$ \\
      \cmidrule{1-5}
      \textit{All-placements} \\
      \cmidrule{1-5}
      CNN1D \cite{Liang2019} & $61.3\pm8.7$ & $58.8\pm9.8$ & $74.0\pm9.0$ & $72.5\pm10.8$ \\
      CNN2D \cite{Ito2018} & $67.5\pm5.9$ & $67.5\pm5.0$ & $76.3\pm5.9$ & $75.0\pm5.3$ \\
      FPbiLSTM \cite{Tang2023} & $75.5\pm5.2$ & $74.0\pm3.7$ & $84.3\pm4.6$ & $81.8\pm3.3$ \\
      DT \cite{reddy2010} & $85.9\pm1.2$ & $76.7\pm2.1$ & $90.1\pm1.2$ & $84.3\pm2.5$ \\
      Fusion-MIL (ours) & $\mathbf{88.1}\pm2.4$ & $\mathbf{86.7}\pm2.1$ & $\mathbf{92.6}\pm1.7$ & $\mathbf{90.8}\pm1.4$ \\
      \bottomrule
    \end{tabular}
  }
\end{table}

To ensure a comprehensive evaluation, our system is compared with prior works
employing both traditional machine learning techniques and deep learning methods
(Table \ref{tab:sota_summary} provides a concise summary of these studies). To
ensure comparability, we replicate the approaches outlined in the original
papers, applying them to the SHL preview Dataset and evaluating their
effectiveness through our experiments. In Table \ref{tab:sota_comparison}, we
present the accuracy and F1-score results of our method against state-of-the-art
approaches, with and without the inclusion of the post-processing step (Hidden
Markov Model). Additionally, we provide the standard deviation across different
sensor placements alongside the scores.

Reddy et al.'s approach \cite{reddy2010} achieves an accuracy of $90.1\%$ in the
all-placements experiment. Consistent with their work, we modify the
classification task by reducing the number of distinct classes to $5$, merging
all $4$ motorized modes into a single class. Their proposed fusion model
incorporates DTs and an HMM with acceleration signals at $32$ Hz and location at
$1$ Hz segmented into $1$-second windows. While our findings reveal very
similar accuracy to what is reported in their paper, a notable deviation in
the F1-Score is observed, likely due to considerable class imbalance in the SHL
preview dataset, accentuated when combining all vehicles into a single category
for the $5$-class task.

Liang et al. \cite{Liang2019} use a lightweight CNN model to recognize $7$ TMs
and achieve accuracy of $61.3\%$ and F1-Score of $58.8\%$. They rely solely on
acceleration signals, sampled at $50$ Hz and segmented into $10$-second
windows. However, it is evident that their method exhibits a significantly lower
performance compared to their reported results. This discrepancy can be
attributed to their practice of using the same participant group for both
training and testing phases. Specifically, they partitioned the dataset into
$80\%$ for training and $20\%$ for testing which most probably influences the
reported outcomes.

Ito et al.'s work \cite{Ito2018} uses FFT spectrograms on $60$-second windows
acquired from acceleration and gyroscope signals data sampled at $100$ Hz. They
train a CNN model with synthesized images produced by arranging two spectrograms
vertically and achieve accuracy of $67.5\%$ and F1-Score of $67.5\%$ in the
all-placements experiment for the task of $8$ TMs.

It is worth noting that \cite{Ito2018} is the only other method of Tables
\ref{tab:sota_summary} and \ref{tab:sota_comparison} that can operate with lower
sampling-rate signals, as it relies on spectrograms. In our experiments, we have
observed no significant loss of effectiveness for lower sampling rates.
However, these effectiveness values are significantly lower than those reported
in the original paper. This deviation can be ascribed to the nature of the
dataset as it was exclusively collected by a single user and was considerably
extensive, amounting to $271$ hours in total with $95$ hours designated for
testing.

Tang et al. \cite{Tang2023} use the same dataset \cite{Ito2018}, demonstrating
enhanced effectiveness through the integration of $3$ distinct sensor types and
the implementation a deep multimodal fusion network. Their approach
combines data from accelerometers, magnetometers, and gyroscopes, each sampled
at $20$ Hz and segmented into $60$-second windows. Introducing a feature pyramid
network, which leverages both shallow-layer richness and deeper-layer feature
resilience, their work achieves accuracy of $75 .5\%$ and F1-Score of
$74\%$ in the all-placements experiment for the task of $8$ TMs.

Notably, there is a clear and consistent ranking in terms of recognition
effectiveness among the various placements of the smartphone: Bag $>$ Hips $>$
Torso $>$ Hand. Furthermore, all methods exhibit a substantial increase in
effectiveness when trained across all different positions, as opposed to the
per-placement experiment where only one position is used for training and
testing. Finally, all methods seem to improve their effectiveness when
incorporating the additional post-processing step (with HMMs).

From the results (Table \ref{tab:sota_comparison}), it can be seen that our
method achieves better effectiveness in all settings and experiments, while
using especially low sampling rates for both the accelerometer and location
data.

\section{Conclusions}
\label{sec:conclusions}

In this work, we have presented an approach to TMR using both acceleration and
location signals. We lower the sampling rate requirements, since this plays a
detrimental role in battery consumption. We proposed a model that extracts
information from each modality separately and then embeds them in a common space
in order to enable the use of an attention-based MIL approach. The final model
is able to automatically use the most suitable modality for each transportation
mode and even cope with missing data from a modality. We evaluate on a publicly
available dataset and achieve high effectiveness (F1 score of $91\%$). We also
evaluate on virtual scenarios where the position of the smartphone changes
placement, and our model is able to maintain its robustness and even increase
its effectiveness, obtaining an F1 score of $92.3\%$. Comparison with literature
algorithms shows better results for our method overall.

\bibliography{IEEEabrv,main}



\begin{IEEEbiography}[{\includegraphics[width=1in,height=1.25in,clip,keepaspectratio]{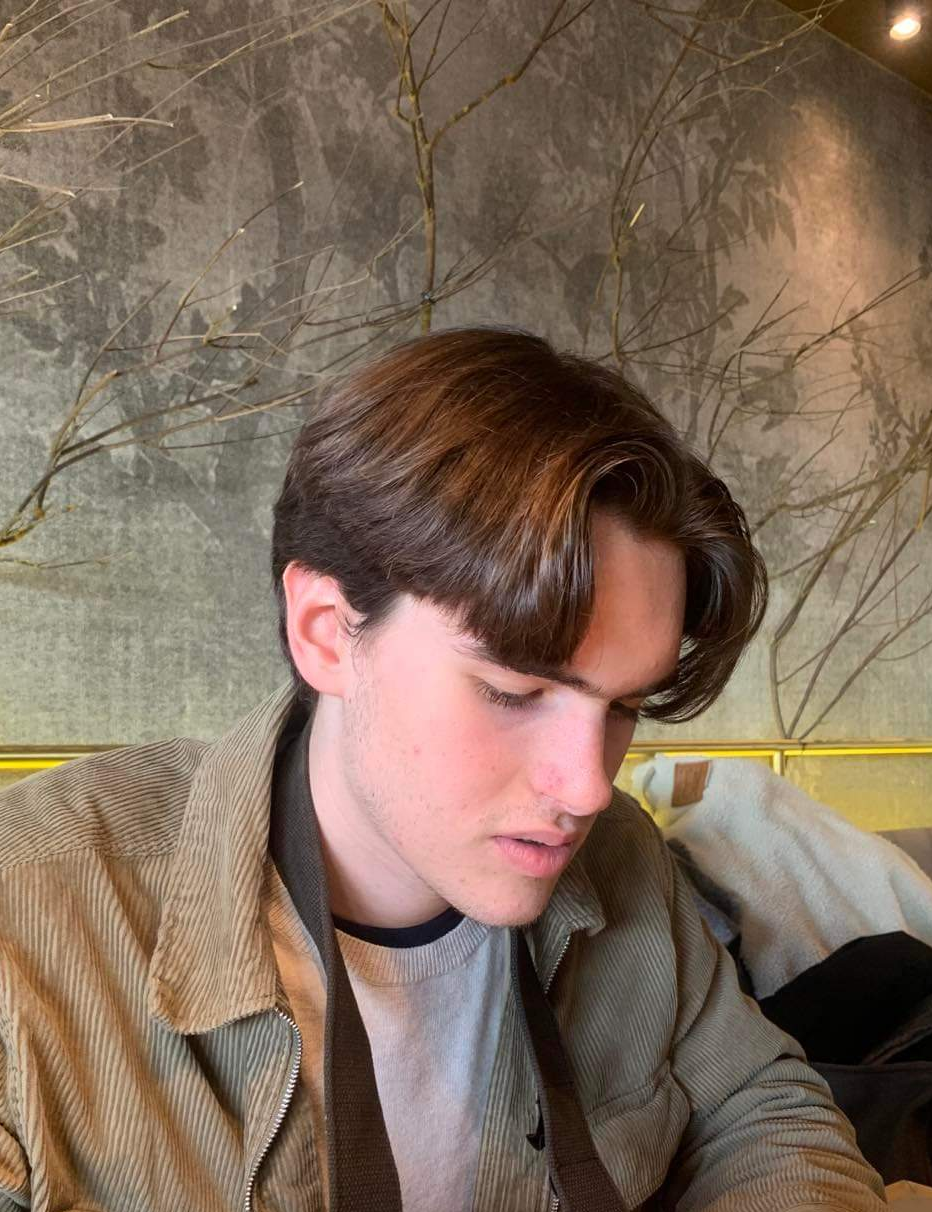}}]
  {Christos Siargkas} is a 4th year undergraduate student at the Department of
  Electrical and Computer Engineering in Aristotle University of
  Thessaloniki. Since October 2022 he has been working as a member of the
  Multimedia Understanding Group, AUTH. His current research interests include
  deep learning with a focus on signal processing.
\end{IEEEbiography}

\begin{IEEEbiography}[{\includegraphics[width=1in,height=1.25in,clip,keepaspectratio]{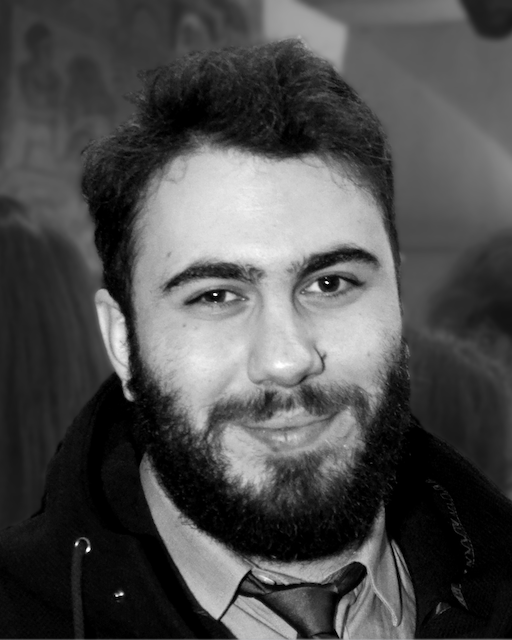}}]
  {Dr. Vasileios Papapanagiotou} is an Assistant Professor at the Department of
  Biosciences and Nutrition of Karolinska Institutet in Stockholm, Sweden. He
  received his Diploma and Ph.D. of Electrical and Computer Engineering in 2013
  and 2019 respectively, both from the Department of Electrical and Computer
  Engineering of Aristotle University of Thessaloniki in Greece. Since 2013 he
  has been mainly working as a researcher and on EU-funded research projects
  (FP7, H2020). His research interests include signal processing and ma- chine
  learning with applications to wearable devices and sensors in the context of
  behav- ioral monitoring and quantification. He is a member of IEEE, IEEE EMBS,
  and the Technical Chamber of Greece.
\end{IEEEbiography}

\begin{IEEEbiography}[{\includegraphics[width=1in,height=1.25in,clip,keepaspectratio]{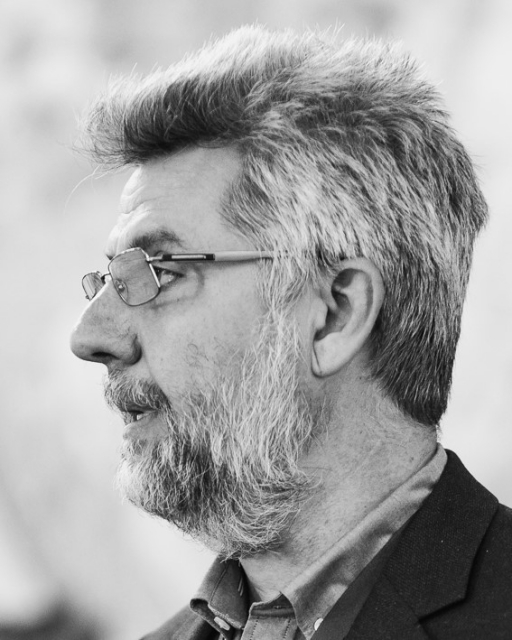}}]
  {Prof.  Anastasios Delopoulos} was born in Athens, Greece, in 1964. He
  graduated from the Department of Electrical Engineering of the National
  Technical University of Athens (NTUA) in 1987, received the M.Sc. from the
  University of Virginia in 1990 and the Ph.D. degree from NTUA in 1993.  From
  1995 till 2001 he was a senior researcher in the Institute of Communication
  and Computer Systems of NTUA. Since 2001 he is with the Electrical and
  Computer Engineering Department of the Aristotle University of Thessaloniki
  where he serves as an associate professor. His research interests lie in the
  areas of semantic analysis of multimedia data and computer vision. He is the
  (co)author of more than 75 journal and conference scientific papers. He has
  participated in 21 European and National R\&D projects related to application
  of signal, image, video and information processing to entertainment, culture,
  education and health sectors. Dr. Delopoulos is a member of the Technical
  Chamber of Greece and the IEEE.
\end{IEEEbiography}

\end{document}